# Unit Commitment with Cost-Oriented Temporal Resolution

Junyi Tao, *Graduate Student Member, IEEE*, Ran Li *Member, IEEE*, Salvador Pineda, *Senior Member, IEEE*

*Abstract*—Time-adaptive unit commitment (UC) has recently been investigated to reduce the scheduling costs by flexibly varying the temporal resolution, which is usually determined by clustering the net load patterns. However, there exists a misalignment between cost and net load patterns due to the discrete start-up costs and out-of-merit-order dispatch triggered by ramping and other constraints. The optimal time-adaptive resolution cannot be completely captured by clustering-based method. This paper proposes a cost-oriented method to address this misalignment by a novel bilevel optimization approach that is efficiently solved through a heuristic greedy algorithm. The impact of varying temporal resolution on the final scheduling costs are tested, based on which the temporal resolution is heuristically updated, achieving significant cost reduction without increasing the number of temporal periods. Subsequently, an improved discretized Adam optimization method together with offline warm start and online refinement strategy is proposed to efficiently search for the better temporal resolution configuration. Results show that the proposed cost-oriented UC temporal resolution determination method achieves enhanced cost efficiency.

## I. INTRODUCTION

THE day-ahead UC determines the on/off status and production levels of all generating units to meet electricity load on an hourly basis [1], [2]. The drawback of this fixed temporal resolution is that during periods of high variability in renewable energy and load patterns, fixed temporal resolution often fails to capture rapid net load changes within the scheduling periods, resulting in significant power imbalance costs [3], [4]. Conversely, during periods of low variability, UC based on fixed temporal resolution yields nearly identical scheduling results [5], [6]. There is a large redundancy of decision variables, leading to a waste of computational resources [7].

Time-adaptive UC has recently been proposed to address these issues [3]. The key of this approach is determining the flexible temporal resolution [6]. Most temporal resolution determination methods are based on the features of the net load curve, with temporal resolution optimized through an objective function minimizing either the sum of absolute gradients [8] or the relative variance range of the net load [9]. Techniques such as agglomerative hierarchical clustering [10] or k-means clustering [3] are also employed to determine the temporal resolution based on the net load curve. Additionally, a sliding window method is proposed to progressively group successive load levels until the average relative error exceeds threshold [11]. Considering that the performance of time-adaptive UC is not only related to the characteristics of the net load but also influenced by generator parameters, a few studies have incorporated the characteristics of power units to determine the temporal resolution. One strategy involves solving approximate solutions of small-scale m-point UC models [12], where the temporal resolution is selected based on extreme values, ramping requirements of the load curve, generator ramping capabilities [13], and changes in generator statuses. Another strategy involves constructing a power output variation model for thermal units [7], with the objective of minimizing the impact on the generation side caused by load side variations.

However, the above clustering-based methods are based on an underlying assumption: the temporal variation of non-cost features (e.g. net load patterns or power unit characteristics) can approximate the variation of real-time scheduling costs. The high penetration of variable renewable energy (VRE) undermines this assumption: (1) The volatility and forecasting error of renewables will frequently violate ramping and other constraints, triggering inconsistent intra-interval start-up costs and out-of-merit-order dispatch [14], [15]. (2) Under high VRE scenarios, scheduling costs in real-time dispatch will become highly volatile [16] and cannot be approximated by non-cost features. These limitations significantly degrade the performance of existing methods in the time-adaptive UC framework. A recent study [17] proposes a performance-driven temporal resolution determination method leveraging well-trained neural networks [18]. The neural networks operate within a closed-loop architecture, allowing feedback on its subsequent UC performance to fine-tune the temporal resolution. However, it only optimizes the number of temporal periods, but the explicit partition is still based on k-means clustering of net loads, which inherit the same problem as discussed above.

This paper proposes a cost-oriented method to adaptively determine the temporal resolution of UC. The impact of varying aggregated time periods on the final real-time scheduling costs is designed as the gain to heuristically update the temporal resolution, seeking for lower costs without increasing the number of time periods. To increase the searching efficiency, an improved discretized Adam optimization method together with offline warm start and online refinement strategy are pro-

This work was supported by the National Natural Science Foundation of China under Grant 52477111. *(Corresponding author: Ran Li).*

Junyi Tao and Ran Li are with the Key Laboratory of Control of Power Transmission and Conversion, Ministry of Education, Shanghai Jiao Tong University, Shanghai 200240, China, and also with the Shanghai Non-Carbon Energy Conversion and Utilization Institute, Shanghai Jiao Tong University, Shanghai 200240, China. Salvador Pineda is with the research group OASYS, University of Malaga, Malaga 29071, Spain (e-mail: taojunyi1004@sjtu.edu.cn; rl272@sjtu.edu.cn; spineda@uma.es).

posed. During periods of high volatility and forecasting error of renewables, where inconsistent intra-interval start-up costs and out-of-merit-order dispatch may occur, the temporal periods in proposed method are subdivided to capture potential variations in scheduling costs. Conversely, the temporal periods are extended to optimize the use of computational resources. Results show that, the proposed cost-oriented method significantly reduces UC scheduling costs. Besides, with the increasing penetration of VRE, the incorporation of probabilistic forecasting enables the proposed method to account for the effect of net load uncertainty, thus resulting in better scheduling results.

## II. Problem Formulation

Conventional UC is typically done by fixing the temporal resolution $x_t^{DA}$ of 1 hour and optimizing over a set of 24 time periods, as shown in Equation (1).

$$x_t^{DA} = h_0, \forall t \quad (1)$$

where $t$ denotes day-ahead temporal period indices, $x_t^{DA}$ denotes the length of the $t$-th day-ahead UC temporal period (h), $h_0$ denotes fixed 1-hour temporal periods (h).

However, since fixed temporal resolution can result in degraded scheduling results, a critical task in this context is determining the better 24 day-ahead temporal periods. The time-adaptive UC method, as referenced in [3], performs clustering on the real-time net load data, which is sampled at a 10-minute resolution, to identify the 24 temporal periods for the day-ahead stage. The initial number of clusters is set to the total number of real-time data points, i.e., 144, and the centroid $\bar{x}_I$ of each cluster $I$ is computed as shown in Equation (2).

$$\bar{x}_I = \frac{1}{S_I} \sum_{i \in I} l_i \quad (2)$$

where $i$ denotes real-time temporal period indices, $I$ denotes the cluster indices, $S_I$ denotes the number of real-time data points in cluster $I$, $l_i$ denotes the real-time net load data in temporal period $i$, $\bar{x}_I$ denotes the centroid of cluster $I$.

The dissimilarity $H(I, J)$ between each pair of adjacent clusters, denoted as $(I, J)$, is then calculated according to Equation (3). The two closest adjacent clusters are merged based on the dissimilarity matrix, continuing this process until the total number of temporal periods is reduced to 24. The day-ahead UC is subsequently performed using the resulting temporal resolution.

$$H(I, J) = \frac{2 S_I S_J}{S_I + S_J} \left\| \bar{x}_I - \bar{x}_J \right\|^2 \quad (3)$$

where $I$ and $J$ denote the adjacent clusters, $S_J$ denotes the number of real-time data points in cluster $J$, $\bar{x}_J$ denotes the centroid of cluster $J$, $H(I, J)$ denotes the dissimilarity between clusters $I$ and $J$.

However, due to the inherent volatility and forecasting error of renewables, real-time dispatch scheduling costs can fluctuate significantly, making them difficult to approximate using non-cost features. For instance, during sunrise and sunset, the rapid fluctuations in PV generation can cause significant short-term decreases or increases in net load. In such cases, the ramping constraints and other constraints can lead to inconsistencies in the changes of net load and scheduling costs. This leads to the misalignment of objectives between minimizing intra-cluster variance of non-cost features and minimizing scheduling costs in real-time dispatch. The misalignment can result in deterioration in scheduling cost performance. Using a typical day in Spanish power system, two critical insights are identified that highlight the significant inconsistency between the net load and scheduling costs.

Insight 1 Start-up cost: As shown in Fig. 1, the black line represents the net load curve, while the red line represents the corresponding scheduling costs. Between 8 a.m. and 9 a.m., a slight increase in net load (from 785 MW to 806 MW) causes a sharp rise in total scheduling cost (from €15,714 to €25,632), whereas, between 9 a.m. and 10 a.m., a small increase of 34 MW in net load results in a significant cost reduction of €7,324. The inconsistency occurs because the output limit of low-cost baseload units is 800 MW. Small variations in net load during this period require the startup of higher-cost intermediate load units, leading to additional scheduling costs.

Insight 2 Out-of-merit-order dispatch: At 20 p.m., a net load of 1,180 MW corresponds to a scheduling cost of €39,031. However, at 22 p.m., a higher net load of 1,220 MW results in a lower scheduling cost of €37,043. The inconsistency arises because, from 19 p.m. to 20 p.m., the net load spikes by 259 MW due to PV reduction. The already-operating base and intermediate load units cannot ramp up quickly enough to meet the increased load due to their ramping constraints, leading to higher-cost peaking units being dispatched to meet the load requirement. Such schedule results in high costs at 20 p.m. (€39,031). However, from 21 p.m. to 22 p.m., the net load decreases by 77 MW. The ramp-down capability of the peaking units allows them to reduce generation, and the remaining power deficit is covered by the intermediate load units, thus resulting in lower scheduling costs at 22 p.m. (€37,043). Therefore, this inconsistency is resulted from the out-of-merit-order dispatch driven by the rapid ramping requirements of net load fluctuation.

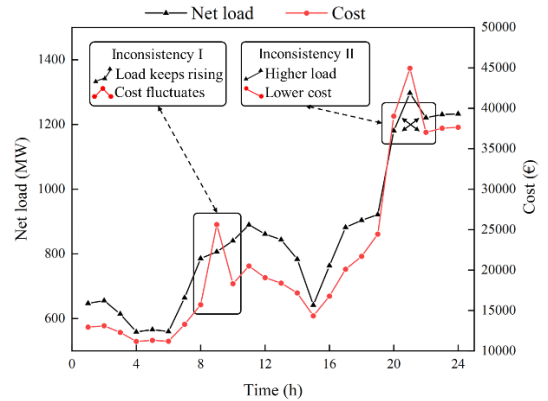

**Fig. 1.** Inconsistency of net load and scheduling cost curve.

These results indicate that, the variability and uncertainty of renewables can lead to inconsistent intra-interval start-up costs and out-of-merit-order dispatch. Significant fluctuations in scheduling costs are thus induced, which cannot be accurately captured by simple non-cost features. Besides, with the increasing penetration of VRE, more frequent start-ups and more rapid ramping requirements of thermal power units will be demanded [19], making the inconsistency between net load and cost more pronounced. The effectiveness of time-adaptive UC based on non-cost features will be greatly undermined. It is necessary to proposed a cost-oriented temporal resolution determination method to align the objective of temporal resolution determination and day-ahead UC with scheduling costs, thereby enhancing the cost efficiency of UC.

## III. COST-ORIENTED TEMPORAL RESOLUTION DETERMINATION METHOD

### A. Rationale of the proposed method

As discussed in Section II, the durations of the coarse temporal periods used in the day-ahead UC stage can significantly impact the scheduling costs obtained from the subsequent real-time economic dispatch (ED) stage, which is performed using fine time resolution. Ideally, the durations of day-ahead temporal periods should be selected in anticipation of their effect on real-time scheduling costs. This can be formulated as a bilevel optimization problem. In upper-level problem, the goal is to determine the durations of the day-ahead temporal periods so as to minimize real-time scheduling costs. The lower-level problem then solves the UC optimization for the given time period configuration, producing generation schedules that serve as inputs to the ED stage. The bilevel structure is illustrated in Fig. 2.

Since the lower-level UC problem involves binary commitment variables, the overall bilevel formulation is computationally challenging. To address this, a heuristic, cost-oriented approach is proposed to efficiently yield high-quality solutions. Specifically, we iteratively and greedily adjust duration of each day-ahead temporal period, selecting at each step the duration that leads to the lowest scheduling costs. This simple yet effective strategy aligns the objectives of time resolution selection and UC optimization, and leads to significant cost reductions.

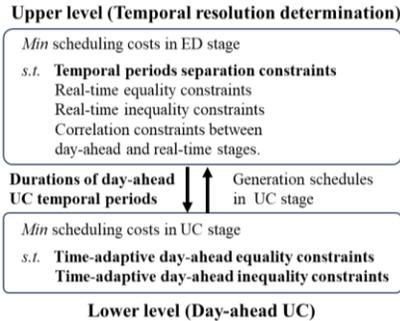

**Fig. 2.** General bilevel process of the proposed method.

Specifically, the day-ahead UC temporal period $x_t^{DA}$ is incorporated as a decision variable in the solving of the optimal scheduling costs problem, which follows a bilevel structure:

**(1) Upper-level:** The upper-level optimization problem determines the temporal resolution in day-ahead UC. As shown in Equation (4), the day-ahead UC temporal resolution function $D()$ is determined with the objective of minimizing real-time scheduling costs. In Equation (5), day-ahead UC temporal period $x_t^{DA}$ is decided by separating the whole length of time periods $L$ (24 h) using the temporal resolution function $D()$, with the objective of minimizing the scheduling costs in real-time dispatch. Equations (6)-(7) define the equality and inequality constraints of real-time dispatch, including power limit constraints, power balance constraints, ramp rate constraints and minimum up/down time constraints. The complete expanded form of Equations (6)-(7) can be found in [3]. Equations (8)-(9) illustrate the correlation between day-ahead decision variables $p_{gt}^{G,DA}$, $u_{gt}^{DA}$ and real-time decision variables $p_{gi}^{G,RT}$, $u_{gi}^{RT}$, reflecting the fact that the commitment of inflexible generation units in the day-ahead stage cannot be modified during the real-time dispatch.

**(2) Lower-level:** The lower-level optimization problem is the day-ahead time-adaptive UC problem. Equation (10) defines the determination of day-ahead decision variables with the objective of minimizing day-ahead UC costs. Equations (11)-(12) represent the equality and inequality constraints of the day-ahead time-adaptive UC. Similarly, the complete expanded form of Equations (11)-(12) can be found in [3].

**Upper level (Temporal resolution determination):**

$$\min_D \sum_{g,i} (C_g^M p_{gi}^{G,RT} x_i^{RT} + s_{gi}^{U,RT} + s_{gi}^{D,RT}) \\ + \sum_i C^{LS} x_i^{RT} (\bar{P}_i^D - p_i^{D,RT}) \quad (4)$$

s.t.

$$x_t^{DA} = D(L) \quad (5)$$

$$F(p_{gi}^{G,RT}, s_{gi}^{U,RT}, s_{gi}^{D,RT}, p_i^{D,RT}, u_{gi}^{RT}) = 0 \quad (6)$$

$$G(p_{gi}^{G,RT}, s_{gi}^{U,RT}, s_{gi}^{D,RT}, p_i^{D,RT}, u_{gi}^{RT}) \leq 0 \quad (7)$$

$$p_{gi}^{G,RT} = p_{gt}^{G,DA}, \forall g \in G_{\text{base}}, \begin{cases} i \in (0, r \cdot x_1^{DA}], \text{ if } t=1 \\ i \in (r \cdot \sum_{h=1}^{t-1} x_h^{DA}, r \cdot \sum_{h=1}^{t} x_h^{DA}], \text{ if } t>1 \end{cases} \quad (8)$$

$$u_{gi}^{RT} = u_{gt}^{DA}, \forall g \in \{G_{\text{base}}, G_{\text{intermediate}}\}, \begin{cases} i \in (0, r \cdot x_1^{DA}], \text{ if } t=1 \\ i \in (r \cdot \sum_{h=1}^{t-1} x_h^{DA}, r \cdot \sum_{h=1}^{t} x_h^{DA}], \text{ if } t>1 \end{cases} \quad (9)$$

**Lower level (Day-ahead UC):**

$$(p_{gt}^{G,DA}, s_{gt}^{U,DA}, s_{gt}^{D,DA}, p_t^{D,DA}, u_{gt}^{DA}) \\ = \arg\min \sum_{g,t} (C_g^M p_{gt}^{G,DA} x_t^{DA} + s_{gt}^{U,DA} + s_{gt}^{D,DA}) \\ + \sum_t C^{LS} x_t^{DA} (\bar{P}_t^D - p_t^{D,DA}) \quad (10)$$

s.t.

$$F'(p_{gt}^{G,DA}, s_{gt}^{U,DA}, s_{gt}^{D,DA}, p_t^{D,DA}, u_{gt}^{DA}) = 0 \quad (11)$$

$$G'(p_{gt}^{G,DA}, s_{gt}^{U,DA}, s_{gt}^{D,DA}, p_t^{D,DA}, u_{gt}^{DA}) \leq 0 \quad (12)$$

where $g$ denotes conventional thermal unit indices, $C_g^M$ denotes marginal production cost of unit $g$ (€/MWh), $C^{LS}$ denotes load shedding cost (€/MWh), $x_i^{RT}$ denotes the length of the $i$-th real-time dispatch temporal period (fixed to 10 minutes), $\bar{P}_i^D / \bar{P}_t^D$ denote the demand at the $i$-th real-time dispatch / the $t$-th day-ahead UC temporal period (MW), $p_{gi}^{G,RT}/p_{gt}^{G,DA}$ denote the power output of unit $g$ at the $i$-th real-time dispatch / the $t$-th day-ahead UC temporal period (MW), $s_{gi}^{U,RT}/s_{gt}^{U,DA}$ denote the start-up cost of unit $g$ at the $i$-th real-time dispatch / the $t$-th day-ahead UC temporal period (€), $s_{gi}^{D,RT}/s_{gt}^{D,DA}$ denote the shut-down cost of unit $g$ at the $i$-th real-time dispatch / the $t$-th day-ahead UC temporal period (€), $p_i^{D,RT}/p_t^{D,DA}$ denote the satisfied demand at the *i*-th real-time dispatch / the *t*-th day-ahead UC temporal period (MW), $u_{gi}^{RT}/u_{gt}^{DA}$ denote the binary variable that equals to 1 if thermal unit *g* is online and 0 otherwise at *i*-th real-time dispatch / the *t*-th day-ahead UC temporal period, *r* denotes the ratio of classic fixed temporal resolution (one hour) to the real-time stage time resolution (10 ten minutes), *L* denotes the total length of temporal periods in day-ahead stage (h), $G_{base}/G_{intermediate}$ denote baseload unit / intermediate load unit sets.

### B. Cost-oriented framework based on greedy algorithm

In Section III-A, the determination process of temporal resolution is elaborated with bilevel structure. However, the bilevel optimization problem is challenging for two main reasons: (1) The lower-level problem includes binary variables $u_{gt}^{DA}$ which denotes the on-off status of generators. As a result, the problem is inherently non-convex and cannot be reformulated using the Karush–Kuhn–Tucker conditions, thus increasing the computational difficulty [20]. (2) The bilevel optimization problem involves complicating constraints (8)-(9), which include both upper-level and lower-level variables, thus increasing the computational scale. Inspired by the core principle of greedy algorithms [21], a cost-oriented framework is proposed to determine the duration of each time period sequentially, from the first to the last. The proposed approach reduces computational effort by decomposing the global optimization problem (4)-(12) into a series of localized subproblems, each involving two neighboring time periods. For each subproblem, the impact of adjusting the day-ahead UC temporal periods on the real-time scheduling costs is evaluated, and this information is subsequently used to update the length of neighboring time periods. The proposed heuristic approach is illustrated in Fig. 3, demonstrating how the greedy-based method determines the temporal resolution.

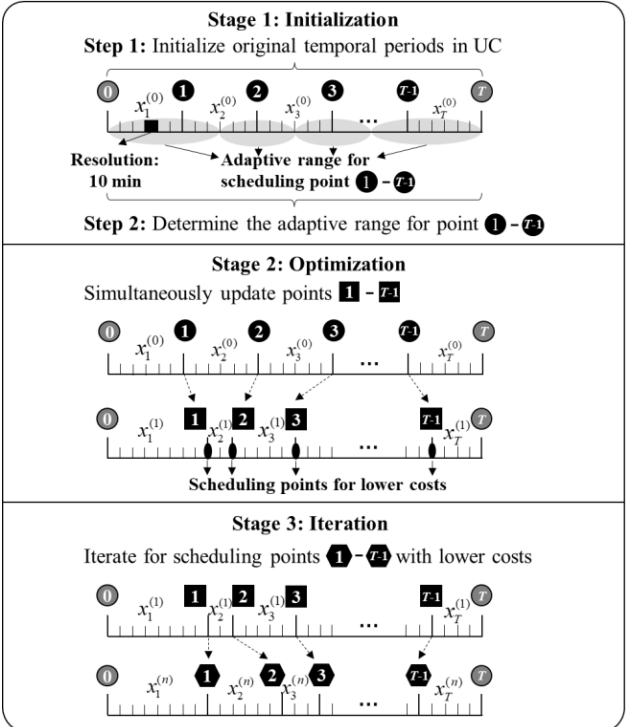

**Fig. 3.** Selection strategy based on greedy algorithm.

### C. Deterministic cost-oriented temporal resolution selection method

To clearly illustrate the proposed temporal resolution selection method, the following assumptions and definitions are provided:

- *In the deterministic scenario*, it is assumed that the net load values for the day-ahead stage are identical to those for the real-time stage, meaning that forecasting errors are not considered.
- *In the probabilistic scenario*, the net load values used are the probabilistic forecast results from 48 hours prior, accounting for the impact of forecasting errors.
- *Scheduling points* refer to the time instances of UC scheduling, denoted by 1, 2, …, *T*.
- *Temporal periods* refer to the duration of UC scheduling intervals, represented by $x_1, x_2, …, x_T$.

**(1) Stage 1: Initialization**. First, the number of day-ahead temporal periods *T* is set to 24 for one day following the practice of traditional UC. It is noted that the method can be applied for any *T* as demonstrated in Section V-C. Then, as shown in Equation (13), an initial length $h_0$ for all *T* temporal periods is set.

$$x_t^{(0)} = h_0, \forall t \tag{13}$$

where $x_t^{(0)}$ denotes the initial length of *t*-th temporal period.

For *T* temporal periods, there are *T*+1 corresponding scheduling points, indexed from 0 to *T*. To maintain a fixed total length of temporal periods, the positions of the first point 0 and last point *T* are held constant, while the positions of the intermediate points 1 to *T*-1 are adjustable. As illustrated in Fig. 3, to achieve full coverage and avoid redundancy, all points are constrained to move only within their adaptive ranges. The maximum and minimum adaptive ranges for an individual scheduling point is determined while keeping the initial positions of the other scheduling points fixed. For scheduling point 2 to *T*-2, the adaptive range encompasses half the distance to the adjacent scheduling point on both the left and right sides. There are two exceptions. For the first scheduling point, the adaptive range includes all available resolutions to its left and half the total distance between itself and the second scheduling point on its right. For the *T*-1-th scheduling point, the adaptive range includes half the total distance between itself and the *T*-2-th scheduling point on its left, as well as all available resolutions to its right. Thus, the maximum and minimum of temporal periods are defined in Equations (14)-(15).

$$x_{t,\max}^{(k)} = \begin{cases} x_t^{(k)} + \dfrac{x_{t+1}^{(k)}}{2}, t < T \\ \dfrac{x_{T-1}^{(k)}}{2} + x_T^{(k)}, \ t = T \end{cases} \tag{14}$$

$$x_{t,\min}^{(k)} = \begin{cases} 0, t = 1, T \\ \dfrac{x_t^{(k)}}{2}, 1 < t < T \end{cases} \tag{15}$$

where *k* denotes the number of iteration times, $x_{t,\max}^{(k)}$ denotes the maximum length of the *t*-th temporal period in the *k*-th iteration, $x_{t,\min}^{(k)}$ denotes the minimum length of the *t*-th temporal period in the *k*-th iteration.

**(2) Stage 2: Optimization.** Given the typical real-time net load data resolution is 10 minutes, we use 10 minutes as the minimum movement interval and move each scheduling point within its feasible range, calculating the day-ahead UC and real-time dispatch results at each position. For the *q*-th scheduling point, the lengths of the *q*-th and *q*+1-th temporal period, $x_q^{DA}$ and $x_{q+1}^{DA}$, are determined by solving the UC optimization problem, as formulated in Equation (16). The constraints include (17)-(19) in upper level and (10)-(12) in lower level. The optimization problems can be solved in parallel for all scheduling points from 1 to *T*-1. It is important to note that since the bilevel optimization problem is decomposed into alternating optimizations at each scheduling point, the remaining temporal periods are treated as fixed when optimizing temporal periods $x_q^{DA}$ and $x_{q+1}^{DA}$, as shown in Equation (19).

**Upper level (Temporal resolution determination):**

$$\min_{(x_q^{DA}, x_{q+1}^{DA})} \sum_{g,i} (C_g^M p_{gi}^{G,RT} x_i^{RT} + s_{gi}^{U,RT} + s_{gi}^{D,RT}) + \sum_i C^{LS} x_i^{RT} (\bar{P}_i^D - p_i^{D,RT}) \quad (16)$$

s.t.

$$x_q^{DA} + x_{q+1}^{DA} = x_q^{(k)} + x_{q+1}^{(k)} \quad (17)$$

$$x_{q,\min}^{(k)} < x_q^{DA} < x_{q,\max}^{(k)} \quad (18)$$

$$x_t^{DA} = x_t^{(k)}, \forall t \neq q, q+1 \quad (19)$$

When adjusting the position of each scheduling point, constraint (17) must be satisfied, ensuring that the total length of the *q*-th and *q*+1-th scheduling periods remains unchanged after optimization, thereby maintaining the overall time period (24 h). Constraint (18) defines the adaptive range of $x_q^{DA}$, and together with constraint (17), the adaptive range of $x_{q+1}^{DA}$ is also defined. Constraint (19) defines the length of the other temporal periods for $t \neq q$ or *q*+1.

The aforementioned alternating optimization problems are solved in parallel for all scheduling points. By targeting the minimal real-time costs through exhaustive search across adaptive range of all scheduling points, the temporal periods $x_q^{DA}$ are determined and stored in $x_{q,alter}^{(k+1)}$, which denotes the length of the *q*-th temporal period in the *k*+1-th iteration in alternating optimization, as shown in Equation (20).

$$x_q^{DA} \rightarrow x_{q,alter}^{(k+1)} \quad (20)$$

The alternating optimization problems are performed for *q*=1 to *T*-1, during which the value of $x_{q,alter}^{(k+1)}$ is obtained based on the condition that the temporal periods except for $x_q^{DA}$ and $x_{q+1}^{DA}$ are treated as fixed. It is necessary to aggregate all $x_{q,alter}^{(k+1)}$ values to get updated temporal periods in the *k*+1-th iteration. The aggregation is carried out as described in (21).

$$x_t^{(k+1)} = \begin{cases} x_{1,alter}^{(k+1)}, t = 1 \\ \sum_{u=1}^{t-1} x_u^{(k)} + x_{t,alter}^{(k+1)} - \sum_{u=1}^{t-1} x_{u,alter}^{(k+1)}, t = 2,...,T-1 \\ L - \sum_{u=1}^{T-1} x_{u,alter}^{(k+1)}, t = T \end{cases} \quad (21)$$

where $x_t^{(k+1)}$ denotes the *t*-th updated temporal period in the *k*+1-th iteration.

**(3) Stage 3: Iteration.** After one iteration, the initial temporal periods are updated to $x_t^{(k+1)}$. An iterative convergence criterion is introduced, as represented in Equation (22), iteration terminates when positions of all points no longer update. If the convergence criterion is not satisfied, and a new round of iteration begins.

$$\sum_{t=1}^{T} \left| x_t^{(k+1)} - x_t^{(k)} \right| = 0 \quad (22)$$

If the convergence criterion is satisfied, this indicates that the bilevel problem is being addressed using a heuristic procedure that yields a feasible, but likely suboptimal solution. Nevertheless, even though the optimal solution of Equations (4)-(12) is not attained, the temporal resolution achieved by the proposed cost-oriented method reduces real-time scheduling costs compared to existing methodologies, as demonstrated by the computational results in Section V.

### D. Probabilistic cost-oriented temporal resolution selection method

As the penetration of VRE increases, the uncertainty of net load will further rise. To address this uncertainty, a promising approach involves leveraging 48 hours prior probabilistic forecasting results to conduct stochastic optimization. Our method exhibits strong applicability under this approach. By incorporating probabilistic forecasting, the cost-oriented temporal resolution determination method can account for net load uncertainty, thereby enhancing the effectiveness of the time-adaptive UC approach. The objective of probabilistic forecasting is formulated in (23), which involves constructing typical scenarios to consider the forecasting error of net load. In Equations (24)-(25), the constraint is rewritten to ensure that the actual output of the units aligns with the actual load demand of each scenario. The remaining Equations are identical to those in the deterministic scenario and are thus not elaborated here.

$$(p_{gt}^{G,DA}, s_{gt}^{U,DA}, s_{gt}^{D,DA}, p_t^{D,DA}, u_{gt}^{DA})$$

$$= \arg\min \frac{1}{S^w} \sum_{w \in S^w} \left\{ \sum_{g,t} (C_g^M p_{gt}^{G,DA} x_t^{DA} + s_{gt}^{U,DA} + s_{gt}^{D,DA}) \right. \quad (23)$$

$$\left. + \sum_t C^{LS} x_t^{DA} (\bar{P}_t^D - p_{t,w}^{D,DA}) \right\}$$

$$\sum_g p_{gt}^{G,DA} = p_{t,w}^{D,DA}, \forall t, \forall w \quad (24)$$

$$0 \leq p_{t,w}^{D,DA} \leq \bar{P}_{t,w}^D, \forall t, \forall w \quad (25)$$

where *w* denotes net load fluctuating scenario indices, $S^w$ denotes the net load fluctuating scenario set.

## IV. ACCELERATION OF COST-ORIENTED METHOD

A key issue with the proposed approach is the computational burden. However, if the minimum temporal resolution is increased to 20 or 30 minutes to release the computational burden, there is a risk of missing the better scheduling results. This section proposes two strategies to accelerate the proposed method: (1) Discrete Adam optimization algorithm and (2) Offline warm start + online refinement.

### A. Discrete Adam optimization algorithm

Inspired by the classical Adam optimization method which is a gradient descent-based algorithm [22], a discrete variant of the Adam algorithm is proposed to improve solution efficiency. This approach transforms the original exhaustive search over discrete points into an adaptive momentum-based learning process, allowing rapid identification of valleys in the cost landscape and yielding low scheduling costs efficiently. In Section III, the optimization of individual scheduling points is described, which can be solved parallelly without temporal dependencies among the results of different scheduling points. Advanced computational tools, such as multi-core parallel processing in MATLAB, can be leveraged in the proposed discrete Adam optimization method to significantly enhance computational efficiency. Specifically, the process of the discrete Adam optimization method is illustrated below.

Initially, the temporal periods are determined by Equation (1) and are fixed at one hour. The minimum time resolution is denoted as $L_{\min}$, which is set to 10 minutes. Subsequently, for the determination of the $q$-th temporal period, the distribution of temporal periods is adjusted by shifting the current scheduling point $q$ to the left by one minimum resolution, as illustrated in Equation (26).

$$\begin{cases} x_{q,left}^{(k+1)} = x_q^{(k)} - L_{\min} \\ x_{q+1,left}^{(k+1)} = x_{q+1}^{(k)} + L_{\min} \end{cases} \quad (26)$$

where $x_{q,left}^{(k+1)}/x_{q+1,left}^{(k+1)}$ represents the length of the $q$-th/$q+1$-th temporal period after scheduling point $q$ is shifted left by one minimum resolution.

The cost $C_{q,left}^{(k+1)}$ is presented in Equation (27). The modified constraints are as shown in Equations (28)-(30), while the remaining constraints are identical to those in Equations (5)-(12).

$$\min \sum_{g,i}(C_g^M p_{gi}^{G,RT} x_i^{RT} + s_{gi}^{U,RT} + s_{gi}^{D,RT}) + \sum_i C^{LS} x_i^{RT}(\bar{P}_i^D - p_i^{D,RT}) \quad (27)$$

$$x_q^{DA} = x_{q,left}^{(k+1)} \quad (28)$$

$$x_{q+1}^{DA} = x_{q,left}^{(k+1)} \quad (29)$$

$$x_t^{DA} = x_t^{(k)}, \forall t \neq q, q+1 \quad (30)$$

where $C_{q,left}^{(k+1)}$ denotes the scheduling cost after scheduling point $q$ is shifted left by one minimum resolution.

Similarly, the distribution of temporal periods is calculated when the current scheduling point is shifted right by one minimum resolution, as shown in Equation (31). The corresponding scheduling cost $C_{q,right}^{(k+1)}$ is calculated using Equation (27), and modified constraints are shown in Equations (32)-(34).

$$\begin{cases} x_{q,right}^{(k+1)} = x_q^{(k)} + L_{\min} \\ x_{q+1,right}^{(k+1)} = x_{q+1}^{(k)} - L_{\min} \end{cases} \quad (31)$$

$$x_q^{DA} = x_{q,right}^{(k+1)} \quad (32)$$

$$x_{q+1}^{DA} = x_{q+1,right}^{(k+1)} \quad (33)$$

$$x_t^{DA} = x_t^{(k)}, \forall t \neq q, q+1 \quad (34)$$

Then, with a minimum time resolution of 10 minutes, we calculate the centered difference of the scheduling cost around scheduling point $q$, as shown in Equation (35).

$$g_q^{(k+1)} = \frac{C_{q,left}^{(k+1)} - C_{q,right}^{(k+1)}}{2 \cdot L_{\min}} \quad (35)$$

For the position determination of the $q$-th scheduling point, given the centered difference $g_q^{(k+1)}$, the corresponding first moment $m_q^{(k+1)}$ and second moment $v_q^{(k+1)}$ are calculated as shown in Equation (36).

$$\begin{cases} m_q^{(k+1)} = \beta_1 \cdot m_q^{(k)} + (1-\beta_1) \cdot g_q^{(k+1)} \\ v_q^{(k+1)} = \beta_2 \cdot v_q^{(k)} + (1-\beta_2) \cdot g_q^{(k+1)\,2} \end{cases} \quad (36)$$

where $m_q^{(k)}$ and $v_q^{(k)}$ denote initial value of first moment and second moment in $k$-th iteration, $\beta_1$ and $\beta_2$ denote parameters.

Adam requires bias correction to eliminate the initial estimation bias, as described in Equation (37).

$$\begin{cases} \hat{m}_q^{(k+1)} = \dfrac{m_q^{(k+1)}}{1-\beta_1^{k+1}} \\ \hat{v}_q^{(k+1)} = \dfrac{v_q^{(k+1)}}{1-\beta_2^{k+1}} \end{cases} \quad (37)$$

where $\hat{m}_q^{(k+1)}$ and $\hat{v}_q^{(k+1)}$ denote bias correction of first moment and second moment in the $k+1$-th iteration.

Finally, we use the corrected moment values to determine the step size $\delta_q^{(k+1)}$ for scheduling point $q$. Given the minimum time resolution is 10 minutes, the resulting step size is adjusted to the nearest multiple of 10 minutes. Besides, the step size must not exceed the maximum adaptive range of scheduling point $q$, as described in Equations (14)-(15). Thus, the determination of step size is illustrated in Equation (38).

$$\delta_q^{(k+1)} = \min\left(\left\lfloor \frac{\frac{\alpha \cdot \hat{m}_q^{(k+1)}}{\sqrt{\hat{v}_q^{(k+1)}}+\varepsilon}}{L_{\min}} \right\rceil \cdot L_{\min}, x_{q,\max}^{(k)}\right) \quad (38)$$

where $\alpha$ is the learning rate, $\varepsilon$ is a small constant added to prevent division by zero ($10^{-8}$), $\lfloor \cdot \rceil$ denotes the rounding function.

Next, after the determination of step size $\delta_q^{(k+1)}$, calculate the updated length $x_{q,alter}^{(k+1)}$ of the $q$-th temporal period in the $k+1$-th iteration, as shown in Equation (39).

$$x_{q,alter}^{(k+1)} = x_q^{(k)} - \delta_q^{(k+1)} \quad (39)$$

Similarly, for all scheduling points $q$ from 1 to $T$-1, the aforementioned discretized Adam optimization method is executed in parallel. By incorporating Equation (21) to aggregate the temporal periods, the temporal periods $x_t^{(k+1)}$ after a single iteration are obtained. Likewise, considering Stage 3: Iteration in Section III-C, the costs are continuously reduced through an iterative approach. The iteration stops when Equation (22) is satisfied, thereby determining temporal periods. This method enhances the solving efficiency of cost-oriented temporal resolution determination method by modifying the traversal process described in Section III into discrete Adam optimization.

*B. Offline warm start and online refinement strategy*

In the proposed methodology, the selection of initial scheduling points position has a significant impact on the iteration times before reaching convergence. The offline warm start and online refinement strategy can effectively reduce the number of iteration times required during the online stage. Specifically, the offline-online process is illustrated below.

The time scales for offline and online methods are defined: The offline period refers to the time window 48 hours prior to the scheduling day. The online period corresponds to the day-ahead scheduling window. Both the offline and online methods employ the previously proposed discrete Adam algorithm, as detailed in Equations (26)-(39), to determine the temporal period results. The key distinction between the two methods lies in the initialization process: the offline method uses initial values derived from a clustering strategy based on net load, while the online method initializes with the converged temporal period results of the offline process.

First, for initializing the offline method, we utilize a net load-based clustering approach. Using high-resolution 48-hour forecasts (10-minute intervals), based on Equations (2)-(3), a hierarchical clustering method is applied to aggregate the temporal periods into 24 segments, which serves as the initial values for the offline method. Then, the cost-oriented strategy combined with the discrete Adam optimization algorithm is applied to iterate and obtain the offline temporal resolution results. Subsequently, these offline temporal resolution results are used as the initial values for the online method. Given the discrepancies between the 48-hour forecast data and actual values, the online method still requires iterative updates to achieve convergence. However, the iteration times is significantly reduced, ensuring that the time requirements for day-ahead and real-time scheduling are met.

## V. CASE STUDY

Using data from the Spanish power system in 2020 [23], this paper validates the effectiveness of the proposed cost-oriented temporal resolution determination method for UC (denoted as CO-UC). A naive cost-oriented approach (NA-UC) is introduced as a comparative case. In the NA-UC approach, the conventional UC with a fixed temporal resolution of 1 hour is first solved. Then, in the ED stage, the scheduling costs are computed with a temporal resolution of 10 minutes. Subsequently, k-means clustering is performed on real-time scheduling costs, merging adjacent time periods with similar costs to determine the flexible temporal resolution for UC. Besides, the CO-UC method is also compared with the conventional UC approach with fixed temporal resolution (denoted as CH-UC), and the time-adaptive UC approach following reference [3] which determines temporal resolution based on net load curve clustering (denoted as TA-UC). Secondly, under uncertainty scenarios, the effectiveness of the CO-UC method combined with probabilistic forecasting is established, showcasing its adaptability to the variability and uncertainty brought by increasing penetration of VRE. Furthermore, the impact of varying the number of temporal periods on the method's performance is evaluated. Finally, the proposed offline warm start and online refinement strategy is tested, highlighting its validation in enhancing computation speed during day-ahead UC and real-time dispatch.

*A. Cost reduction in deterministic scenario*

To evaluate the effectiveness of the proposed CO-UC method, a case study was conducted using data over January 2020. Table I presents the start-stop times and scheduling costs resulting from ED stage under different temporal resolution determination methods of UC. The CO-UC method reduces unit start-stop times to 6.60 per day, resulting in a corresponding reduction in start-up and shut-down costs to €15,093. This represents a notable improvement over the suboptimal TA-UC method, which incurred 8.70 start-stop times per day and €19,740 in associated costs. Moreover, when compared to the other methods, the CO-UC approach achieves the highest proportion of base load units (79.84%) and the lowest proportion of peaking units (2.83%), resulting in a lower scheduling cost overall. These results demonstrate that the CO-UC approach achieves the most efficient merit-order dispatch of generation units. Ultimately, the total cost of the proposed CO-UC method per day is €1,309,723, demonstrates an average reduction of 0.73% in total costs compared to the suboptimal method (TA-UC) and an average reduction of 2.11% compared to the worst-performing method (CH-UC), thereby showcasing better cost reduction performance.

To analyze the cost reduction mechanisms of the proposed CO-UC method compared to TA-UC method, Fig. 4(a) presents the correlation of net load and cost in TA-UC at 10-minute temporal resolution during real-time ED stage in January 16. Two segments, denoted as (I) and (II), are selected to highlight the inconsistencies between net load and cost. In inconsistency I, the net load remains relatively stable with slight changes, whereas the cost exhibits significant fluctuations, with notable price spikes of €7,819 occurring at 4:30 a.m. and 5:10 a.m.. In inconsistency II, the net load exhibits sharp upward then downward trends, and the cost does not follow the same trend but instead shows several pronounced spikes, such as €10,178 at 10:40 a.m. and €8,810 at 11:30 a.m., both higher than costs in adjacent periods. These results demonstrate the evident inconsistencies between net load and cost during real-time dispatch in Spanish power system.

TABLE I
DAILY AVERAGE COST REDUCTION OF THE PROPOSED CO-UC APPROACH IN JANUARY 2020 IN DETERMINISTIC SCENARIO

| Method | Start-stop times | Start-stop cost (€) | Baseload unit operation cost (€) | Intermediate load unit operation cost (€) | Peaking unit operation cost (€) | Total Cost (€) | Cost reduction percentage |
|---|---|---|---|---|---|---|---|
| CH-UC | 11.33 | 25,127 | 1,034,746 (78.83%) | **228,268 (17.39%)** | 49,647 (3.78%) | 1,337,886 | |
| NA-UC | 8.63 | 19,593 | 1,033,735 (79.00%) | 226,464 (17.31%) | 48,275 (3.69%) | 1,328,303 | -0.72% |
| TA-UC | 8.70 | 19,740 | 1,033,933 (79.57%) | 223,858 (17.23%) | 41,606 (3.20%) | 1,319,371 | -1.38% |
| CO-UC | **6.60** | **15,093** | **1,033,550 (79.84%)** | 224,412 (17.34%) | **36,582 (2.83%)** | **1,309,723** | **-2.11%** |

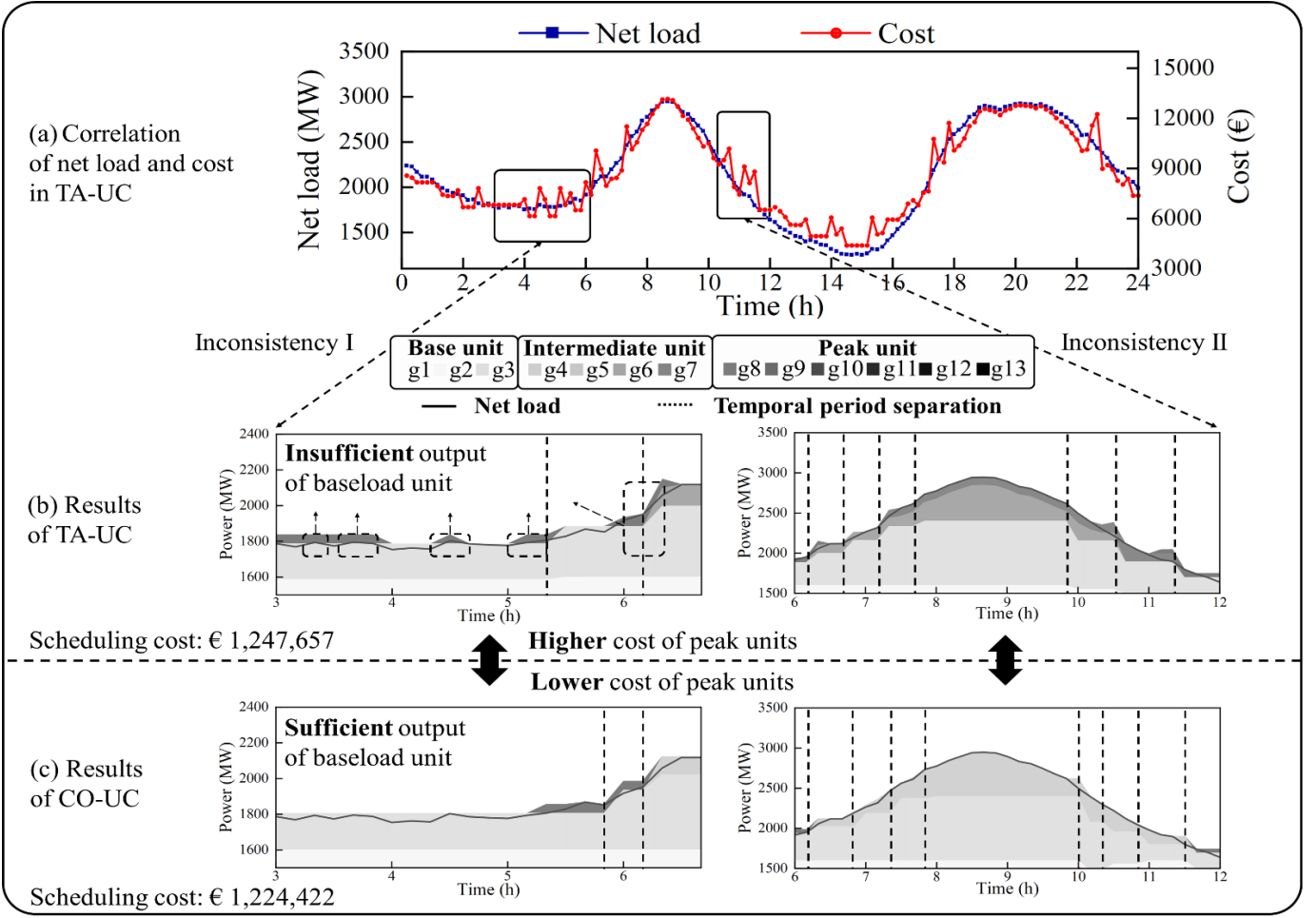

**Fig. 4.** Detailed analysis of cost reduction in proposed CO-UC approach.

Fig. 4(b) and Fig. 4(c) present the real-time ED results of TA-UC and CO-UC methods. The vertical dashed lines indicate the segmentation of temporal periods. The first inconsistency represents a typical period of low net load variability (e.g., 3:00-6:40 a.m.). During this stage, the TA-UC method, which clusters periods based on net load patterns, tends to aggregate low-variability periods to reduce computational burden. However, as discussed in Section II of this paper (Insight 1), due to unit output limit constraints, when a single unit operates at its maximum output, even small variations in net load during this period necessitate the startup of higher-cost units, leading to additional startup and operation costs. The TA-UC method is unable to mitigate the additional startup costs caused by these constraints, leading to frequent instances of insufficient baseload unit output, as shown in the left panel of Fig. 4(b), and thus resulting in higher costs for peaking units, as observed in inconsistency 1 of Fig. 4(a). In contrast, the proposed CO-UC method can guide the selection of day-ahead temporal periods based on real-time cost variations through the closed-loop algorithm, even during periods of stable load fluctuations. As shown in the left panel of Fig. 4(c), even though the total number of temporal periods is kept the same as in the TA-UC method, the CO-UC approach explicitly accounts for the impact of minor net load changes on real-time startup costs. This ensures sufficient baseload unit output and significantly reduces the start-stop times of peak units, thereby effectively lowering the overall scheduling costs.

The second inconsistency encompasses periods characterized by rapid increases and decreases in load (e.g., 6:00-12:00 a.m.). By comparing the dashed lines in the right panels of Fig. 4(b) and Fig. 4(c), it is evident that during this period, the CO-UC method yields a more detailed segmentation of temporal periods compared to the TA-UC method. The TA-UC method utilizes more intermediate and peaking units which have higher costs to meet the real-time net load demand, whereas the CO-UC method makes more efficient use of lower-cost baseload units. The reason for this difference in generation output mix lies in the fact that, while the TA-UC method effectively meets load demands through net load clustering, it does not account for the ramping up and down constraints of different generation types. The temporal resolution is often unfavorable for baseload units, which have stringent ramping constraints. As a result, it frequently requires the startup and shutdown of peaking units, which have more relaxed ramping constraints, to meet load demands. In contrast, the proposed CO-UC method facilitates a more cost-effective approach to meeting load demands by maximizing the utilization of baseload units with lower operating costs, thereby reducing reliance on peaking units for ramping, startup and shutdown. Ultimately, this results in significant cost reductions. The scheduling cost un-der the TA-UC method is €1,247,657, while under the CO-UC method the cost is €1,224,422, achieving a cost reduction of 1.86%.

*B. Cost reduction in probabilistic scenario*

The effectiveness of incorporating probabilistic forecasting into the temporal resolution determination methods is evaluated in this Section. When traditional clustering methods based on load or cost are applied to probabilistic scenarios, it becomes challenging to identify the most likely load or cost curves. A straightforward approach is to perform temporal periods aggregation based on the average load curves or costs across these scenarios. With the net load fluctuations become more severe, rendering clustering based on average loads or costs across different scenarios will be unrepresentative in guiding the selection of temporal periods for UC. In contrast, for the probabilistic forecasts of the net load, the proposed CO-UC method first applies stochastic optimization, as described in Section III-D, to calculate the expected costs across all forecasting scenarios. This expectation is then used to determine the optimal day-ahead UC temporal resolution, still utilizing the heuristic approach illustrated in Fig. 3. The applicability of the proposed method under probabilistic scenarios was evaluated in Table 2. The results indicate that, the NA-UC and TA-UC approaches exhibit similar cost performance, each yielding a 0.93% cost reduction compared to CH-UC. In comparison, the CO-UC method provides an additional 1.20% cost reduction over the NA-UC and TA-UC approaches, resulting in a total reduction of 2.12% compared to CH-UC. The CO-UC approach reduces costs in probabilistic scenario significantly.

TABLE 2
COST REDUCTION IN PROBABILISTIC SCENARIO

| Method | Total cost (€) | △$C_{suboptimal}$ | △$C_{worst}$ |
| --- | --- | --- | --- |
| CH-UC | 1,305,204 | 0.93% | 0% |
| NA-UC | 1,293,103 | 0% | -0.93% |
| TA-UC | 1,293,118 | 0% | -0.93% |
| **CO-UC** | **1,277,564** | **-1.20%** | **-2.12%** |

*C. Cost reduction in different number of temporal periods*

To investigate the applicability of the proposed method on different number of temporal periods, four methods are compared with the number of temporal periods ranges from 1 to 24 as shown in Fig. 5. The blue, green, gray, and red lines represent CH-UC, NA-UC, TA-UC, and CO-UC, respectively. Fig. 5 demonstrates that, as the number of aggregated temporal periods increases, scheduling costs generally exhibit a gradual decreasing trend. Specifically, compared to the other methods, the proposed CO-UC method consistently achieves lowest scheduling costs across all aggregated temporal resolutions, indicating its strong robustness. Comparing to the suboptimal approach (TA-UC), CO-UC achieves the maximum cost reduction of 2.91% when the number of aggregated periods is three. The average cost reduction across all numbers of aggregated periods is 1.41%. The number of aggregated temporal periods in the CO-UC method can be flexibly adjusted based on actual scheduling requirements to achieve a trade-off between scheduling costs and the overall number of unit dispatches, ultimately attaining optimal benefits.

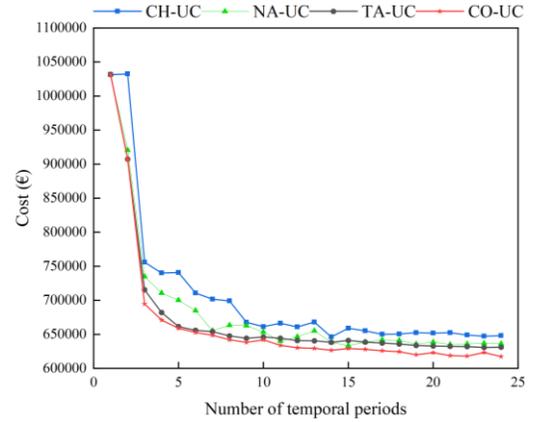

**Fig. 5.** Costs with different number of temporal periods.

*D. Acceleration with warm start and refinement strategy*

To investigate the warm-start approach proposed in this study, forecast results from 48 hours prior are used to determine the initial temporal resolution during the warm-start stage. For comparison, a scenario without the warm start method was considered, where the CO-UC approach is directly performed day-ahead. The scheduling costs of CO-UC with and without warm start throughout the iterative process are illustrated in the Fig. 6. It is observed that when employing the offline warm start scheme, convergence is achieved by the second iteration of the discrete Adam optimization, and real-time scheduling costs are significantly reduced to €1,096,537 compared to the starting point €1,136,119 of CH-UC. In contrast, without the offline warm start, the initial temporal periods for day-ahead UC are determined using the TA-UC method, achieving convergence only by the ninth iteration, with scheduling costs from the first to eighth iteration remaining relatively high. These results demonstrate the effectiveness of the proposed offline warm start combined with online refinement method in accelerating the temporal resolution determination process for UC.

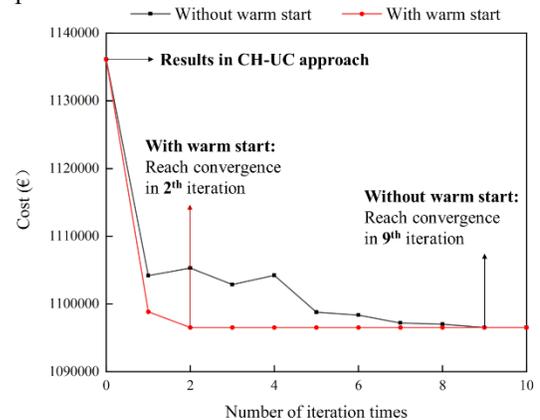

**Fig. 6.** Iteration of CO-UC with and without warm start.

Table 3 presents the computation time of temporal resolution determination in different methods, as well as the time for solving the UC and ED problems over one day. For each method, since the number of time periods remains consistent, the computational time for UC and ED is approximately the same, around 5 seconds. Regarding the time required for determining the temporal resolution, CH-UC adopts fixed temporal resolution, incurring no additional computation time. NA-UC first solves the UC and ED with fixed temporal resolution and then performs clustering based on scheduling costs, resulting in a temporal resolution determination time of 5.24 seconds. In contrast, TA-UC straightly performs clustering on net load, only requiring 0.53 seconds. For CO-UC, the computation time is 49.82 seconds with offline warm start and 477.50 seconds without, both achieving the same cost reduction. The above tests are conducted on a 10-core computational platform. If sufficient cores are deployed to enable high-performance parallel computing, all the alternating optimization processes can be executed concurrently. In that case, the total computation time for CO-UC with offline warm start is theoretically three times that of CH-UC, making it feasible for practical applications.

TABLE 3
COMPUTATION TIME OF DIFFERENT METHODS

| Method | Temporal resolution determination time (s) | UC and ED solving time (s) |
| --- | --- | --- |
| CH-UC | 0 | 4.52 |
| NA-UC | 5.24 | 4.77 |
| TA-UC | 0.53 | 4.83 |
| CO-UC without warm start | 477.50 | 4.66 |
| CO-UC with warm start | **49.82** | 4.75 |

VI. CONCLUSION

This paper presents a cost-oriented method for determining the temporal resolution of day-ahead UC. Compared to existing clustering-based strategies, the proposed CO-UC method can guide the selection of day-ahead temporal periods based on real-time cost variations through a bilevel optimization model that is efficiently solved by a heuristic greedy algorithm. The results demonstrate that, by capturing the high startup and shutdown costs of units caused by ramping and other constraints, the method optimizes temporal resolution. The resulting generation schedule of the proposed method maximizes the dispatch of baseload units with low generation costs, reducing the dispatch of peaking units for start-ups and ramping operations, and ultimately lowering total scheduling costs. The proposed method exhibits strong effectiveness to aggregate temporal periods since it properly accounts for the impact of net load fluctuations on costs both in the deterministic and probabilistic case. To further enhance the application of the proposed method, the introduced UC acceleration technique effectively reduces the number of iterations and improves computational efficiency, thereby satisfying real-time scheduling requirements.

Future research will focus on further enhancing the computational efficiency of CO-UC method. For instance, integrating advanced decision-focused learning techniques into the CO-UC framework to develop end-to-end machine learning models for temporal resolution determination [24].